\documentclass{aastex}
\usepackage{spr-astr-addons}
\usepackage{url}
\usepackage{color}

\RequirePackage{color}

\begin{document}

\title{Electromagnetic Fields and Charged Particle Motion Around
Magnetized Wormholes}
\slugcomment{}
\shorttitle{}
\shortauthors{Ahmedov et al.}

\author{A.A. Abdujabbarov\altaffilmark{1,2}} \and \author{B.J.Ahmedov\altaffilmark{1,2,3}}


\altaffiltext{1}{Ulugh Begh Astronomical Institute,
Astronomicheskaya 33, Tashkent 100052, Uzbekistan}
\altaffiltext{2}{Institute of Nuclear Physics,
        Ulughbek, Tashkent 100214, Uzbekistan}
\altaffiltext{3}{International Centre for Theoretical Physics,
Strada Costiera 11, 34014 Trieste, Italy}

\begin{abstract}
\noindent We perform a study to describe motion of charged
particles under the influence of electromagnetic and gravitational
fields of a slowly rotating wormhole with nonvanishing magnetic
moment. We present analytic expression for potentials of
electromagnetic field for an axially symmetric slowly rotating
magnetized wormholes. While addressing important issues regarding
the subject, we compare our results of motion around black holes
and wormholes in terms of the ratio of radii of event horizons of
a black hole and of the throat of a wormhole. It is shown that
both radial and circular motions of test bodies in the vicinity of
a magnetized wormhole could give rise to a peculiar observational
astrophysical phenomenon.
\end{abstract}

\keywords{Magnetized wormholes \and Electromagnetic fields \and
Charged particle motion}

\section{Introduction}
\label{intro}

Astrophysical objects called wormholes link widely separated
regions of a Universe or of two different Universes, joining two
different spacetimes \cite{mt88,visser95}.

The subject of strong electromagnetic fields due to highly
magnetized rotating neutron stars like pulsars and magnetars is of
great relevance for the  physics of a Wormhole (WH) and for the
particle motion around it, specially around its throat.

{In paper \cite{teo} author in details considered solution for
rotating WH and properly described ergoregion, which surrounds the
throat at the equator of WH.}

A wormhole may be a reason for gravitational lensing
effects~\cite{Dey} and may also create a black hole through
accretion of matter~\cite{Kardashev}. Thus, it is  a subject of
astrophysical importance. The motion of particles around a
wormhole and a possible dragging of such moving particles towards
its vicinity  constitute a subject of Physical reality. To make a
Lorentzian wormhole traversable and stable, one uses exotic
matter, which violates the well-known energy conditions according
to the need of the geometrical structure~\cite{mtyu88, naresh1,
naresh2}. Viable models for such a WH have recently been studied
\cite{Kuhfittig08, lobo1, lobo2, sushkov, lobo3, lobo4}.

The rotation of a magnetized star in vacuum induces electric
field~\cite{d55}. General Relativiity (GR) generates additional
electric field (See, for example, \cite{mt92,kk00,ram01,ram01b})
through its role in the context of dragging of inertial frames and
becomes very important in pulsar magnetosphere~\cite{bes90,mt90}.
Under the framework of GR, slowly rotating wormholes have been a
subject of study, particularly in the context of stress-energy
tensor~\cite{Perez}, scalar fileds~\cite{Kashargin,Kim} and
electromagnetic fields~\cite{Jamil}. The exact solutions of the
wormhole with classical, minimally coupled, massless scalar field,
and electric charge are discussed in the paper \cite{kim}. They
concluded that the addition of electric charge might change the
gravitational field of the WH but will not change the spacetime
seriously.

Here, we focus on the  motion of charged test particles in
gravitational and electromagnetic field of slowly rotating
wormhole with magnetic dipole momentum. We use Hamilton-Jacobi
equation to find influence of {both} the fields {on} the effective
potential due to radial motion of test particles. In
section~\ref{potential}, we calculate potential of electromagnetic
field due to axially-symmetric slowly rotating magnetized
wormhole.

We then consider the separation of variables in the
Hamilton-Jacobi equation and derive the effective potential for
the motion of charged particles around slowly rotating wormhole
with dipolar electromagnetic field in
Section~\ref{motion_charged}. We also calculate stable circular
orbits for charged particles in terms of the magnetic moment of
the wormhole and offer a table. {We present the numerical results
for periods of anharmonic oscillations of charged particles}.
Finally, we conclude our main findings {and present some
astrophysical applications of our results} in
section~\ref{conclusions}. Our present investigation of the motion
of charged particles around a slowly rotating magnetized wormhole
involving these potentials is carried out with the aim to find
astrophysical evidence for the existence of such objects and to
explore its possible differences with the other class objects
called black holes. {Finally, on the contrary to the model
\cite{Kardashev}, for our WH model which has a magnetic dipole
momentum one can obtain the observable difference on circular
motion of charged particle around WH and around compact object as
stars, BH etc.}

Throughout the paper, we use a space-like signature $(-,+,+,+)$
and a system of units in which $G = 1 = c$ (However, for those
expressions with an astrophysical application we have written the
speed of light explicitly.). Greek indices are taken to run from 0
to 3 and Latin indices from 1 to 3; covariant derivatives are
denoted with a semi-colon and partial derivatives with a comma.

\section{Potential of the Electromagnetic Field Around  a Wormhole}
\label{potential}

We may safely ignore quadratic term of the angular velocity
($\omega$) of the free falling frame because of the slow rotation
of the wormhole. Thus, the metric that describes spacetime around
an axialily symmetric slowly rotating wormhole, may be written in
the following form ~\cite{shatskiy1,shatskiy2}:
\begin{eqnarray}\label{metric}
&& ds^{2}=-e^{2\phi(r)}\cdot
dt^{2}+\left[1-\frac{b(r)}{r}\right]^{-1}
dr^{2}\nonumber\\
&&\quad+r^{2}\left(d\theta^{2}+ \sin^{2}\theta
d\varphi^{2}\right)-2\omega(r)r^2\sin^{2}\theta d\varphi dt.
\end{eqnarray}
Here, $\emph{r}$  is the radial coordinate, $\phi(\emph{r})$  is
the so-called lapse function, $b(\emph{r})$  is the shape
function. The $\omega(r)=2J/r^{3}$ is also known as Lense-Thirring
angular velocity, where $J$ is the total angular moment of the
gravitating object. The neck of the wormhole corresponds to the
minimum ${r}={r}_{0}=b({r}_{0})$, where we have $\partial
b/\partial r|_{r_{o}}\leq 1$. The presence of a horizon implies
$\phi\rightarrow-\infty$ or $e^{\phi}\rightarrow0$ such that
$\phi$ is finite everywhere.

Solution of the Eintein equations for WH has been compared with
Reissner-Nordtstrom solution for compact objects with upper limit
for magnetic charge in ref.~\cite{shatskiy1}, wherein components
of the WH metric (\ref{metric}) have been written as
\begin{equation} \label{1stcomp}
 \exp \phi=\left(1-\frac{r_{h}}{r}\right)^{1+\delta}\ ,
\end{equation}
and
\begin{equation}\label{2ndcomp}
b(r)=r_{h}\left[1+\left(1- \frac{r_{h}} {r}\right)^{1-
\delta}\right]\ .
\end{equation}
The quantity $\delta$ in the above expressions, (\ref{1stcomp})
and (\ref{2ndcomp}),  may be found from transcendental equation
$b(r_{0})=r_{0}$:
\begin{equation}\label{delta}
\delta=\frac{\ln\left (\frac{r_{h}}{r_{0}}\right)}
{\ln\left(1-\frac{r_{h}}{r_{0}}\right)}.
\end{equation}

Common to considered  model by \cite{shatskiy1} is the assumption
that the tunnel of magnetic WH is penetrated by an initial
magnetic field, which should display a radial structure for an
external observer in the spherically symmetrical case; i.e., it
should be correspond to a macroscopic magnetic monopole. It means
that metric (\ref{metric}) with expressions (\ref{1stcomp}) and
(\ref{2ndcomp}) correspond to the wormhole mode of the substance
which is the monopole magnetic field and dipole electric field in
slowly rotating approximation (equations (12) and (25) of the
paper \cite{shatskiy1}):
\begin{eqnarray}
\label{B_monopolar}
&& {B}^{\rm \hat{r}}\approx \frac{q_{m}}{r^2} \ ,\\
\label{E1_monopolar} && {E}^{\rm \hat{r}}\approx
\frac{2ar_h^2}{r^3}B^{\rm \hat{r}}(r_h)
\cos\theta \ ,\\
\label{E2_monopolar} && {E}^{\rm \hat{\theta}}\approx
\frac{ar_h^2}{r^3} B^{\rm \hat{r}}(r_h)\sin \theta \ ,
\end{eqnarray}
where $q_{m}$ can be interpreted as magnetic monopole of magnetic
WH described with metric (\ref{metric}) (in addition with
expressions (\ref{1stcomp}) and (\ref{2ndcomp})). Here $\hat $
(hat) stands for orthonormal components of the electric and the
magnetic fields:
\begin{equation}
\label{definition}
E_{\alpha}=F_{\alpha\beta}u^{\beta},\,\,\,B^{\alpha}=-\frac12
\eta^{\alpha\beta \gamma\rho}F_{\beta\gamma} u_{\rho}\ ,
\end{equation}
that are measured by zero angular momentum observer (ZAMO) with
four velocity
$$u^{\alpha}=e^{-\phi}(1,0,0,\omega),\,\,\,
u_{\alpha}=e^{\phi}(1,0,0,0),$$
where
$$
\eta^{\alpha\beta \gamma\rho}=-\frac{1}{\sqrt{-g}}
\epsilon_{\alpha\beta\gamma\rho},
$$
$\epsilon_{\alpha\beta\gamma\rho}$ is the Levi-Civita symbol.

Tensor of electromagnetic field in the presence of monopole
magnetic charge $q_{m}$ can be described as \cite{cabibbo}:
\begin{eqnarray}
\label{cabibbo1} &&
F_{\mu\nu}=A_{\nu,\mu}-A_{\mu,\nu}-\eta_{\mu\nu\rho\sigma}
\tilde{A}^{\sigma,\rho}\ ,\\
\label{cabibbo2} &&
\tilde{F}_{\mu\nu}=\tilde{A}_{\nu,\mu}-\tilde{A}_{\mu,\nu}+
\eta_{\mu\nu\rho\sigma} {A}^{\sigma,\rho}\ ,
\end{eqnarray}
which is included not only the 4-vector potential $A_\mu $, but
also a pseudovector $\tilde{A}_\mu$.
 Lagrangian of the system can be written as :
\begin{eqnarray}
&& L=L_{\rm em}+L_{\rm int}+L_{m}=\nonumber\\
&& - \frac{1}{2} [n^{\mu}(A_{\nu,\mu}-A_{\mu,\nu})]^{2} -
\frac{1}{2} [n^{\mu}(\tilde{A}_{\nu,\mu}-\tilde{A}_{\mu,\nu})]^{2}
\nonumber\\
&&-\frac{1}{4}
\varepsilon_{\mu\nu\rho\sigma}n^{\nu}n^{\nu}\tilde{A}^{\sigma,\rho}
n^{\alpha}A^{\mu}_{\ ,\alpha}\nonumber\\
&&+ \frac{1}{4} \varepsilon_{\mu\nu\rho\sigma}
n^{\nu}n^{\nu}{A}^{\sigma,\rho} n^{\alpha}
\tilde{A}^{\mu}_{\,\alpha}\ ,
\end{eqnarray}
where $n^{\mu}$ is an arbitrary fixed unit four-vector.
Hamilton-Jacobi equation corresponding to this system can be
written as:
\begin{eqnarray}
g^{\alpha\beta}\left(\frac{\partial S}{\partial
x^{\alpha}}+eq_{\rm m} \tilde{A}_{\alpha}\right)
\left(\frac{\partial S}{\partial x^{\beta}}+eq_{\rm m}
\tilde{A}_{\beta}\right)=-m^2.
\end{eqnarray}
The value of pseudovector $\tilde{A}_\mu$ being responsible for
the electromagnetic field
(\ref{B_monopolar})--(\ref{E2_monopolar}) can be found using
equations (\ref{definition}) and (\ref{cabibbo1}),
(\ref{cabibbo2}) as
\begin{equation}
\tilde{A}^\alpha=\frac{q_{m}}{r}\left(0,0,0,\frac{1-\cos
\theta}{\sin\theta}\right)\ ,
\end{equation}
having singularity at $\theta=\pi$ which can be removed by
coordinate transformations. 

Using the variable separation technique one can easily find
equation of radial motion of charged particle around magnetic WH
possessed macroscopic magnetic monopole:
\begin{eqnarray}
f(r, \delta)\left(\frac{dr}{d\sigma}\right)^2={\cal E}^2-V_{\rm
eff}(q_m, r, \tilde{{\cal L}})\ ,
\end{eqnarray}
where $\tilde{{\cal L}}\Rightarrow {\cal L}+eq_m$, ${\cal L}$ is
the angular momentum of the charged particle moving around
magnetic monopole. $f(r, \delta)$ is a function which tends to
$1/2$ when  spacetime metric (\ref{metric}) becomes flat one and
\begin{equation}
V_{\rm eff}=\frac{\tilde{{\cal L}}^2}{2m r^2}\ .
\end{equation}

Study of the influence of magnetic monopole to the motion of
charged particles around WH can be not considered when the massive
WH monopolar magnetic field of the magnetized WH is negligible
($B\lesssim 10^{7} G$ for object with total mass $M\backsimeq
10^{6}M_{\bigodot}$, see table 1 and equation (12) of the paper
\cite{shatskiy1}, with compare to the dipolar magnetic field as
$B\sim 10^{12} G$ of magnetized object. This magnetic field can be
created by the stellar azimuthal electric currents). It is not
difficult to show that the electromagnetic corrections created by
the magnetic dipole being proportional to the electromagnetic
energy density are rather small in most WHs. Indeed if $\rho$ is
the average rest mass density off WH of total mass $M$ and radius
of the throat $r_h$ as measured at infinity, these corrections are
at most
\begin{eqnarray}
&& \hspace{-0.5cm}\frac{B^2}{8\pi \rho_0 c^2}\simeq \\
&& 6.7\cdot10^{-3} \left(\frac{B}{10^{12}\ {\rm G}}\right)^2
\left(\frac{10^6 M_{\bigodot}}{M}\right)
\left(\frac{r_h}{2\cdot10^{6}\ {\rm km}}\right)^3 .\nonumber
\end{eqnarray}

The influence of electromagnetic field with WH monopolar
configuration on the test particles motion should be taken in
account for lower mass WHs, which can be studied in is the future
investigations.

We may now consider the general form of Maxwell's equations
written as:
\begin{eqnarray}
&& 3!F_{[\alpha\beta ,\gamma]}=2(F_{\alpha\beta
,\gamma}+F_{\gamma\alpha ,\beta}+ F_{\beta\gamma
,\alpha})=0\,,\label{maxeq1}\\ && {F^{\alpha\beta}}_{;\beta}=4\pi
J^{\alpha},  \label{maxeq2}
\end{eqnarray}
where $F_{\alpha\beta}=A_{\beta,\alpha}-A_{\alpha,\beta}$ is the
electromagnetic field tensor, $A_\alpha$ is the four potential of
the electromagnetic field and  $J^{\alpha}$ is the four-electric
current.

{Next, we describe a few assumptions that are going to be used
hereafter. First, we assume there is no matter outside the WH so
that the conductivity $\sigma=0$ for outside. We also assume that
the magnetic moment of the WH does not vary in time by supposing
very high conductivity of the WH matter, where magnetic field
produced. However the components of the electromagnetic field will
change periodically due to misalignment between the direction of
magnetic dipole $\mathbf{\mu}$ and axis of rotation}

In the presence of the the magnetic dipole momentum of the
wormhole, the four potential has two nonvanishing components only:
\begin{eqnarray}
&& \label{0thpot} A_{0}=\frac{\mu\Omega
r_{0}^{2}}{3r^3}\left[\cos\chi \left (
3\cos^2\theta-1\right)\right. \nonumber\\
&&\hspace{1cm} \left. + 3\sin\chi \cos\lambda \sin\theta
\cos\theta\right]\,,\\
&& \label{3rdpot} A_{3}=\frac{2\mu}{r}\left(\cos\chi \sin\theta -
\sin\chi \cos\theta \cos\lambda \right) \ ,
\end{eqnarray}
according to the expressions for the four potentials derived in
paper~\cite{ram01} for slowly rotating magnetized neutron star.
For the solutions (\ref{0thpot})-(\ref{3rdpot}) the exact external
solutions of the Maxwell equations (\ref{maxeq1})-(\ref{maxeq2}),
take the following form~\cite{ram01}:
\begin{eqnarray}
&& E^{\hat{r}}= -\frac{\mu\Omega
r_0^2e^{-\phi}}{r^4}\sqrt{1-\frac{b(r)} {r}}\nonumber\\
&&\hspace{1cm} \times \left[\cos\chi\left(3\cos^2\theta-1
+\frac{8M}{5r}
\sin\theta\right)\right.\nonumber\\
&&\hspace{1cm} \left. +\sin\chi\cos\lambda\left(\frac{3}{2}\sin
2\theta -\frac{8M}{5r} \cos\theta \right)\right]\,,\label{Errhat}\\
&& E^{\hat\theta}=\frac{2\mu\Omega r_0^2
e^{-\phi}}{r^4}\left[\cos\chi\cos\theta\left(\sin\theta+ \frac{4M}
{5r} \right)\right. \nonumber\\
&&\hspace{1cm} \left.+\sin\chi\cos\lambda\left(\cos
2\theta+\frac{4M}{5r} \sin\theta\right) \right]\,,\label{Etthat}\\
&& E^{\hat\varphi}=-\frac{\mu\Omega e^{-\phi}}{r^2}
\left(\frac{r_0^2}{r^2}+2\csc\theta\right)\sin\chi
\sin\lambda \cos\theta\ ,\label{Ephihat}\\
 && B^{\hat{r}}=\frac{2\mu}{r^{3}}\left(\sin\chi
 \cos\lambda + \cos \chi
\cot\theta\right)\,,\label{Brrhat}\\
&& B^{\hat{\theta}}=\frac{2\mu}{r^{3}} \sqrt{1-\frac{b(r)} {r}}
\left( \sin\chi \cot\theta \cos\lambda -\cos \chi \right)\,,\label{Btthat}\\
&& B^{\hat{\varphi}}=\frac{2\mu}{r^{3}} \sin\chi \sin\lambda
\cot\theta \ , \label{Bphihat}
\end{eqnarray}
$\mu$ is the magnetic moment of the wormhole, $\Omega$ is the
angular velocity, $M$ is the total mass (see, for example,
~\cite{shatskiy1}), $\chi$ is the inclination angle of the
magnetic moment relative to the rotation axis and
$\lambda(t)=\varphi-\Omega t$ is the instantaneous azimuthal
position.

\section{Motion of the Charged Particles Around Slowly Rotating
Magnetized Wormhole} \label{motion_charged}


The Hamilton-Jacobi equation
\begin{equation}\label{HJ1}
g^{\mu\nu}\left(\frac{\partial S}{\partial
x^\mu}+eA_\mu\right)\left(\frac{\partial S}{\partial
x^\nu}+eA_\nu\right)=m^2\ ,
\end{equation}
may be used as a tool in investigating the motion of charged
particles only when separation of variables may be put to effect.

It was demonstrated by~\cite{dt02} that for metrics describing
axially symmetric spacetimes, variables in Hamilton-Jacobi
equations may be made separated provided action $S$ is separable
as
\begin{equation}\label{action}
S=-{\cal E}t+{\cal L}\varphi+S_{\rm r\theta}(r,\theta)\ .
\end{equation}

Using expressions, (\ref{0thpot}) and (\ref{2ndcomp}), equation
(\ref{HJ1}) may be written in the following forrm:
\begin{eqnarray}
 &&
 \left(\bigg\{{\cal E}-
\frac{e\mu\Omega r_{0}^{2}}{3r^3}\left[\cos\chi \left(
3\cos^2\theta-1\right) + \frac{3}{2}\sin\chi
\cos\lambda\right.\right.\nonumber\\
&& \left.\left. \times \sin2\theta \right]\bigg\}^2+
\frac{8Mr_{0}^{2}\Omega {\cal E}}{5
r^3}\right)\left(1-\frac{r_{h}}{r}\right)^{-2(1+\delta)} \nonumber
\\
&&  +\left\{1-\frac{r_{h}}{r}\left[1+\left(1-\frac{r_{h}} {r}
\right)^{1-\delta} \right]
\right\}\left(\frac{\partial S_{\rm r\theta}}{\partial r} \right)^2\nonumber\\
&& +\frac{1}{r^2}\left(\frac{\partial S_{\rm r\theta}}{\partial
\theta} \right)^2 +\frac{1}{r^2 \sin^2 \theta} \times\bigg[{\cal
L}+ e\frac{2\mu}{r}\nonumber \\
&&\left(\cos\chi \sin\theta - \sin\chi \cos\theta
\cos\lambda\right) \bigg]^2=m^2. \label{HJeq}
\end{eqnarray}
It is not possible to separate variables in this equation for  the
general case but it may be made possible for the motion in the
equatorial plane $\theta=\pi/2$. However this is not allowed as
there are no particle orbits under the Lorentz force confined to
the equatorial plane. For this reason we are forced hereafter to
choose the inclination angle of the magnetic moment relative to
the rotation axis as $\chi=0$ in order to keep particles in
equatorial plane. Indeed under this condition electromagnetic
field of wormhole forces charged particle to move in the
equatorial plane. This can be seen from solutions
(\ref{Errhat})-(\ref{Bphihat}): in the equatorial plane the
$\hat{E}^\theta$ component of the electric field disappears and
only $\hat{B}^\theta$ component of the magnetic field is
nonvanishing. Then the equation for radial motion of charged
particles takes the form
\begin{equation}\label{radmot}
\left(\frac{dr}{d\sigma}\right)^2={\cal E}^2-V_{eff}^2\left(r,
\mu, \Omega, r_{0}, r_{h}, \delta, {\cal E}, {\cal L} \right)\ ,
\end{equation}
where quantity
\begin{eqnarray}\label{effpotrm}
&& V_{eff}^2=\left[1-\frac{r_{h}}{r}-\frac{r_{h}}{r}\left(1-
\frac{r_{h}}{r}\right)^{1-\delta}\right]\nonumber\\
&& \times\left[\frac{1}{r^2}\left({\cal L}+\frac{2e \mu}
{r}\right)^2 -1-\left(1-\frac{r_{h}}{r}\right)^{-2(1+\delta)}
\right.\nonumber\\
&& \left. \times\left\{{\cal E}^2+\frac{r_{0}^2 \Omega{\cal
E}}{15r^4} \big[ (5r+24M) 2e\mu- 8Mr{\cal L}\big]\right\}
\right]\nonumber\\
\end{eqnarray}
can be thought of as effective potential of the radial motion of
charged test particle, where $\sigma$ is the proper time along the
trajectory of the particle, and we change quantities as  $\ {\cal
E} \rightarrow {\cal E}/m$, $\ {\cal L} \rightarrow {\cal L}/m$,
and $\ \mu \rightarrow \mu/m$.

\begin{figure}

  a)    \includegraphics[width=0.45\textwidth]{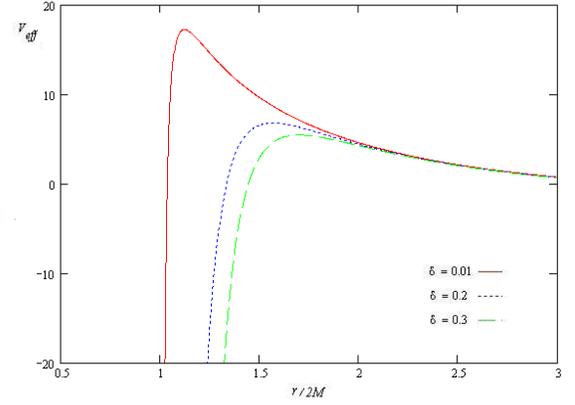}

  b)    \includegraphics[width=0.45\textwidth]{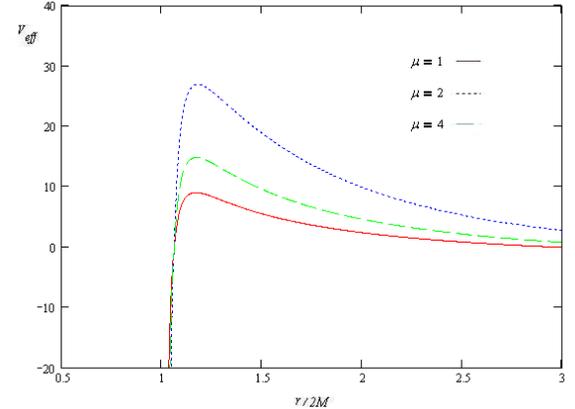}

\caption{{Radial dependence of the effective potential of radial
motion of charged particle near the magnetized WH (a) for
different values of the parameter $\delta$ and (b) for different
values of the magnetic dipole momenta $\mu$. }}
\label{fig:1}       
\end{figure}

{Figure \ref{fig:1} shows the radial dependence of the effective
potential of the radial motion of the charged test particle in the
equatorial plane of the slowly rotating magnetized WH for
different value of the parameter $\delta$ (a) and the magnetic
dipole momenta $\mu$ (b). From this dependence one can obtain
radial motion of charged particle in the equatorial plane of the
WH . As it is seen from the figure \ref{fig:1} the parameter
$\delta$ changes the shape of effective potentials near the
object. In the case of far distances from central object influence
of parameter is negligible, which means that one can see the
difference between WH and black hole (or compact object with
non-exotic matter) only near these objects.}

{Motion of charged particle in the presence of this kind of
effective potential can be explained as follows: increasing of the
magnitude of magnetic dipole momenta of the WH may make circular
objects to be more unstable and let particle go away to infinity.
From the potential we can infer the qualitative structure of the
particles orbits. As it is seen from the figure \ref{fig:1} the
potential carries the repulsive character. It means that the
particle coming from infinity and passing by the source will not
be captured: it will be reflected and will go to infinity again as
it was in the case of black holes. For weak electromagnetic field
of the WH particles can follow bound orbits depending on their
energy. As magnetic dipole momenta $\mu$ increases following
feature arises: the orbits start to be only parabolic or
hyperbolic and no more circular or elliptical orbits exist.}

{From the equation (\ref{HJeq}) one can easily get equations
describing the motion of the test particle what is done below.}

{Trajectory of the charged particle around slowly rotating
magnetized WH can be drawn from the following equation:}
{\begin{eqnarray}
\label{traj} && \left(\frac{dr}{d\varphi}\right)^2
=\nonumber\\
&& \frac{\left\{r_h r^2\left[1+\left(1-\frac{r_h}{r} \right)^{1-
\delta}\right]-r^2\right\}
\left(1-\frac{r_h}{r}\right)^{-2(1+\delta)}}{\left\{{\cal
L}+\frac{2e\mu}{r}-r^2 ({\cal E}+ \frac{e\mu\Omega
r_{0}^{2}}{3r^3})
(1-\frac{r_h}{r})^{-2(1+\delta)}\omega\right\}^2}\nonumber\\
&&  \times\left\{\left[ \left({\cal
L}+\frac{2e\mu}{r}\right)^2-r^2 \right]
\left(1-\frac{r_h}{r} \right)^{2(1+\delta)}-r^2 \right. \nonumber\\
&& \left.\times\left({\cal E} +\frac{e\mu\Omega
r_{0}^{2}}{3r^3}\right) \left(\frac{e\mu\Omega
r_{0}^{2}}{3r^3}+\frac{4e\mu\omega}{r}+{\cal E}+2{\cal L} \omega
\right)\right\}. \nonumber\\
\end{eqnarray} }

\begin{figure}
  \includegraphics[width=0.45\textwidth]{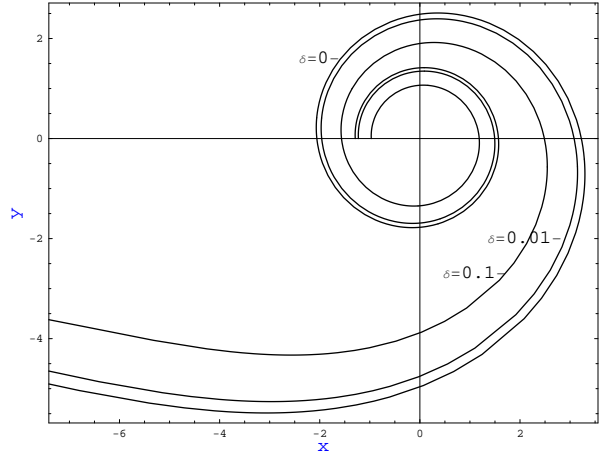}

\caption{{Shape of the trajectory of charged particles around
magnetized WH for the different value of the parameter $\delta$.
}}
\label{fig:2}       
\end{figure}

{It is almost impossible to integrate equation (\ref{traj}) in
general form. However one can get shape of the trajectory of the
test particle by using basic assumptions and numerical
integration. Figure \ref{fig:2} illustrates the shape of the
trajectory of charged particles starting from sufficiently far
distances towards the slowly rotating central object for different
values of small parameter $\delta$ and zero momenta of the
particle in infinity. From the presented figure \ref{fig:2} one
can see that increase of the parameter $\delta$ makes
gravitational field of the central object more stronger which
forces test particle approach closer to the central object.}

{Radial motion of the charged particles near the slowly rotating
magnetized WH can be described using the following equation
derived from (\ref{HJeq}):}
{\begin{eqnarray}
\label{rad} &&
\left(\frac{dr}{dt}\right)^2=\nonumber\\
&&\frac{\left\{r_h\left[1+\left(1-\frac{r_h}
{r}\right)^{1-\delta}\right]-1\right\}\left(1-\frac{r_h}{r}\right)^{2
(1+\delta)}} {\left\{r\left(\frac{e\mu\Omega
r_{0}^{2}}{3r^3}+\frac{2e\mu\omega}{r} \right)+
r\left[E+L\omega\right]\right\}^2}
\nonumber\\
&& \times\left\{\left[ ({\cal L}+\frac{2e\mu}{r})^2-r^2 \right]
\left(1-\frac{r_h}{r} \right)^{2(1+\delta)} -r^2\right. \nonumber\\
&& \left.\times\left({\cal E} +\frac{e\mu\Omega
r_{0}^{2}}{3r^3}\right) \left(\frac{e\mu\Omega
r_{0}^{2}}{3r^3}+\frac{4e\mu\omega}{r}+{\cal E}+2{\cal L} \omega
\right)\right\}.\nonumber\\
\end{eqnarray} }

{In paper \cite{novikov} it was shown from the solution of
equation of radial motion of particle in spherical symmetric
spacetime of nonmagnetized WH that particles can make radial
harmonic oscillations. We obtain here from equation (\ref{rad}) in
the case of magnetized slowly rotating WH, that charged particles
make radial anharmonic oscillations. The periods of that
oscillations are presented in the Table 1. for the different
values of the magnetic parameter and the parameter $\delta$}.
\begin{table*}\label{tab1}
\caption{Period of the radial oscillating of charged particles
near the slowly rotating magnetized WH with respect to $\mu$ and
$\delta$. }
\begin{center}
\begin{tabular}{llllll}
\hline\noalign{\smallskip}
$\delta$   & 0.001   & 0.01    &  0.02   &  0.05   &  0.1   \\
\noalign{\smallskip}\hline\noalign{\smallskip}
$\mu=4 $   & 4.02782 & 4.07698 & 4.11094 & 4.21451 & 4.39281 \\
$\mu=6 $   & 4.41272 & 4.46564 & 4.52357 & 4.69716 & 4.9919  \\
$\mu=12$   & 4.93646 & 4.99437 & 5.05869 & 5.25325 & 5.58609 \\
$\mu=20$   & 5.22254 & 5.28301 & 5.35083 & 5.55727 & 5.91227 \\
\noalign{\smallskip}\hline
\end{tabular}
\end{center}
\end{table*}

Finally we study periods of the circulating charged particle
around slowly rotating magnetized WH (stability of the circular
orbits will be discussed later in the next subsections) by using
the following equation derived from (\ref{HJeq}):
\begin{eqnarray}
&&\left(\frac{d\varphi}{dt}\right)^2=\nonumber\\
&&\frac{\left\{({\cal L} +\frac{2e\mu}{r})
(1-\frac{r_h}{r})^{2(1+\delta)}-r^2 ({\cal E} +\frac{e\mu\Omega
r_{0}^{2}}{3r^3})
\omega(r)\right\}^2}{r^4\left\{\left(\frac{e\mu\Omega
r_{0}^{2}}{3r^3}+{\cal E} \right)+ \left[\frac{2e\mu}{r} + {\cal
L}\right]\omega(r)\right\}^2}.\nonumber\\
\end{eqnarray}

\begin{figure}
  \includegraphics[width=0.45\textwidth]{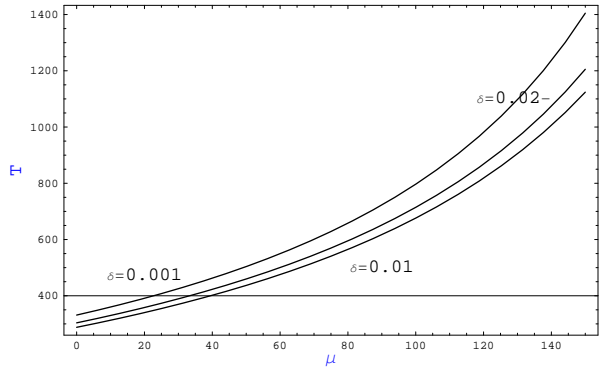}

\caption{{Dependence of the period of the motion of charged
particles around WH from magnetic dipole momenta of central object
for the different values of the parameter $\delta$.}}
\label{fig:3}       
\end{figure}

{Figure \ref{fig:3} shows the dependence of the period of the
circulating particle from the magnetic dipole momenta of the WH
for different values of the small parameter $\delta$. The graphs
justify that increase of the parameter $\delta$ cause particles to
approach closer to the central object.}

\subsection{Stable Circular Orbits for Charged Particles }\label{subs_circ_orb}
   %
Special interest for the accretion theory of test particles around
a slowly rotating wormhole with a dipolar electromagnetic field is
related to the study of circular orbits which are possible in the
equatorial plane $\theta=\pi/2$ when $dr/d\sigma$ is zero.
Consequently the right hand side of equation (\ref{radmot})
vanishes:
\begin{equation}\label{1cond}
{\cal E}^2-V_{eff}^2\left(r, \mu, \Omega,, r_{0}, r_{h}, \delta,
{\cal E}, {\cal L} \right)=0\ ,
\end{equation}
along with its first derivative with respect to r
\begin{equation}\label{2cond}
\frac{\partial V_{eff}}{\partial r}=0\ .
\end{equation}
The radius of marginal stability, the associated energy and
angular momentum of the circular orbits may be derived from the
simultaneous solution of the condition:
\begin{equation}\label{3cond}
\frac{\partial^2 V_{eff}}{\partial r^2}=0\ .
\end{equation}
From equations (\ref{1cond}) and (\ref{2cond}), one may find
expression for energy
\begin{eqnarray}\label{expenrgy}
&& {\cal E}=e\gamma C_0(1-t)^3 \pm \nonumber\\
&& \left(\big[e\gamma C_0(1-t)^3+ \omega r_{h}\kappa(1\pm
\alpha)\big]^2-\gamma \big[ 1- \right.\nonumber\\
&& \left. (1-t)^2\kappa^{2}{\left(1\pm \alpha \right)}^2\big]
\right)^{1/2}+ \omega r_{h}\kappa\left(1\pm \alpha \right)
\end{eqnarray}
and expression for momentum
\begin{equation}\label{expmomenta}
{\cal L}=-eC_{3}(1-t)+ r_{h}\kappa\left(1\pm \alpha\right)\
\end{equation}
of charged test particles. Here we have used the following
notations:

$$ \alpha=\left(1-\frac{\beta r_{h}}{\kappa C_3}\right)^{1/2}\ ,
\qquad \kappa=\frac{C_{3}\ln t}{1+\ln t-3t -6t\ln t}\ ,$$
$$
\beta=\frac{(1+2\ln t)t-1-\ln t}{\ln t} \ , \qquad
C_3=\frac{2e\mu}{r_{h}}\ ,
$$
and
$$
\gamma^{-1}=4t^2\delta \ln t\ .
$$
Now, inserting  (\ref{expenrgy}) and (\ref{expmomenta}) into
equation (\ref{3cond}), one may obtain the basic equation
\begin{eqnarray}\label{radeqtn}
&& {\cal L}r^3\bigg[-4C_3 \lambda+{\cal
L}(r-7\lambda)\frac{r}{r_{h}}\bigg]\nonumber\\
&&+r\bigg[\frac{2C_1}{r_{h}}\lambda^2
 +\frac{r^5}{r_{h}}(3\lambda-r )-4\delta {\cal E}\lambda^2
  \left({\cal E}r^3-2\eta {\cal L } \right)\bigg] \nonumber \\
&&- 2\lambda \bigg[2C_3 \bigg\{4\delta {\cal E} \eta
 \lambda^2+{\cal L} (r-4\lambda
 )\frac{r^3}{r_{h}}\bigg\}-C_3^2\lambda
  r^2 \nonumber\\
&& +
\frac{r}{r_{h}}\bigg\{C_1\frac{13\lambda-5r}{r_{h}}\lambda-r^6+{\cal
L}^2\frac{6\lambda-3r}{r_{h}}r^3\nonumber\\
&&+2\delta {\cal E}\lambda\big[{\cal E}(5r-7 \lambda)r^3 +2\eta
{\cal L}(13\lambda-5r)\big]\bigg\}\bigg]
 \ln\frac{\lambda}{r}\nonumber\\
&& +4\delta {\cal E}\lambda^2
r_{h}\left[2C_3\eta \lambda\frac{4\lambda-3r}{r_{h}^2}\right.\nonumber\\
&& \left.  +3r\left({\cal E} r^3\frac{2\lambda-r}{r_{h}^2}+C_0
\varsigma +2\eta {\cal L} \varsigma\right)\right]
\ln^2\frac{\lambda}{r}=0.
\end{eqnarray}
In this,  we have used following additional notations:
$$
r-r_{h}=\lambda\ ,\  \eta=4Mr_{0}^2\Omega/5 \ ,\ C_{0}=-e\mu\Omega
r_0^2/3r_{h}^3
$$
$$
\varsigma=7r_{h}^2-8r_{h}r+2r^2 \   {\rm and} \quad
C_1=2C_0\delta{\cal E}r_{h}^4.
$$
%


The numerical solutions of the equation (\ref{radeqtn}) will
determine the radii of marginally stable circular orbits for
slowly rotating wormholes with  magnetic dipole momenta as
functions of the parameter $\delta$, the angular velocity
$\Omega$, as well as of the magnetic dipole $\mu$ of the source.
In the Table \ref{tab2}, we  numerical solutions for the radii of
the stable circular orbits of charged test particles for different
values of the parameter $\delta$ and magnetic dipole moment $\mu$
of the wormhole. With the increase of the $\delta$, radii of
stable circular orbits shift to observer at the infinity, while
existence of magnetic dipole moment and a possible increase in it
displace the orbits to the gravitational source.

\begin{table*}\label{tab2}
\caption{Radii of the stable circular orbits of test particles
near the slowly rotating magnetized wormhole with respect to $\mu$
and $\delta$. }
\begin{center}
\begin{tabular}{llllll}
\hline\noalign{\smallskip}
$\delta$ & 0.01 & 0.02 &  0.03 &  0.04 &  0.05   \\
\noalign{\smallskip}\hline\noalign{\smallskip}
$\mu=0.3$ & 7.81538 & 9.31114 & 10.3082 & 11.0739 & 11.7021 \\
$\mu=0.7$ & 6.31054 & 7.52564 & 8.34287 & 8.97572 & 9.49905 \\
$\mu=2$   & 4.86001 & 5.79180 & 6.41939 & 6.90621 & 7.30947 \\
$\mu=5$   & 3.90299 & 4.64910 & 5.15093 & 5.53995 & 5.86208 \\
$\mu=13$  & 3.18162 & 3.79127 & 4.19940 & 4.51495 & 4.77577 \\
$\mu=21$  & 2.93889 & 3.50249 & 3.87818 & 4.16793 & 4.40702 \\
\noalign{\smallskip}\hline
\end{tabular}
\end{center}
\end{table*}

\section{Astrophysical Applications and Conclusions}
\label{conclusions}

{In this paper we consider electromagnetic field of slowly
rotating WH i.e. neglecting quadratic and higher order terms of
angular velocity we first find exact vacuum solutions of Maxwell
equations in spacetime of slowly rotating magnetized WH. In the
paper ~\cite{Kim} it has been justified that electric charge could
exist in WH and did not change the structure of the space-time
around WH seriously, which gives us the right to consider the
existence of magnetic dipole momenta of the WH due to possible
motion of the electric charge.}

{Astrophysical} processes and effects around black hole and WH
different models are distinguishable such as absence of event
horizon on WH, passage of the radiation and particles through WH,
appearance of blueshift effect in addition to the gravitational
redshift effect near WH etc (see for the details
\cite{Kardashev,shatskiy2}). In Ref. \cite{shatskiy1} circular
orbits of test particles around a WH and their periods were
studied. Here we extended these results to motion of charged
particles and have shown the strong dependence of particle motion
from WH shape parameter $\delta$ and magnetic field strength.

{Here we address two basic question: are there any differences
between astrophysical processes around WH and standard models for
compact object? If there are so, how can we distinguish WH from
other compact objects using observational data? To answer these
questions we consider magnetic dipole momentum of the WH and
develop the existing WH model constructed by~\cite{shatskiy1}. For
such models we perform a charged particle motion analysis around
magnetized WH. Finally we can conclude that from astrophysical
point of view the following differences between WH and standard
compact objects can be detected observationally:}

1. Oscillations of bodies in the vicinity of a WH throat (radial
orbits) could give rise to a peculiar observational phenomenon.
Signals from such sources detected by an external observer will
display a characteristic periodicity in their spectra. All objects
(stars, BHs) other than WHs absorb bodies falling onto them
irrecoverably. Periodic radial oscillations are a characteristic
feature of magnetized WHs as it was first shown by
\cite{Kardashev} for other WH models.

2. As it was shown by \cite{Kardashev} the difference of the
circular motion of particles around  Reisner–Nordstr\"{o}m black
hole  and around the charged WH model described by
\cite{shatskiy1} is negligible at $r = 2r_{h}$ (or more). Thus all
conclusions about circular orbit around a WH are the same as in
the limiting Reisner–Nordstr\"{o}m geometry at the corresponding
distances. However, on the contrary for our WH model which has a
magnetic dipole momentum one can easily obtain the difference on
circular motion of charged particle around WH and compact object
as stars, BH etc. {As we have shown in subsection
\ref{subs_circ_orb} with the increase of the $\delta$, radii of
stable circular orbits shift from central object (WH) to observer
at the infinity. From the numerical results given in Table 2. one
can easily see that the influence of $\delta$ parameter on
particle motion depends on the strength of magnetic field of WH:
radii of the stable circular orbits differ almost 2 times bigger
when $\mu=21$, while this increasing almost disappears when
magnetic field of WH is margin.}

\section*{Acknowledgments}

AAA and BJA thank the IUCAA for warm hospitality during their stay
in Pune and AS-ICTP for the travel support through BIPTUN (NET-53)
program. Authors also thank Dr. Anisul Usmani for useful comments
and correcting the text of paper. This research is supported in
part by the UzFFR (projects 5-08 and 29-08) and projects
FA-F2-F079, FA-F2-F061 of the UzAS and by the ICTP through the
OEA-PRJ-29 project and the Regular Associateship grant.

\bibliographystyle{spr-mp-nameyear-cnd}

\end{document}